\documentclass[12pt,english,onecolumn, draftcls, conference]{IEEEtran}

\usepackage[T1]{fontenc}
\usepackage[latin1]{inputenc}
\usepackage{geometry}
\usepackage{amsthm}
\usepackage{amsmath}
\usepackage{amssymb}
\usepackage{graphicx}
\usepackage{epstopdf}
\usepackage{cite}
\usepackage{color}
\usepackage{mathtools}
\usepackage{cases}
\usepackage{stackengine}
\usepackage{accents}

\newcommand\blfootnote[1]{%
  \begingroup
  \renewcommand\thefootnote{}\footnote{#1}%
  \addtocounter{footnote}{-1}%
  \endgroup
}

\newtheorem{thm}{Theorem}

\newtheorem{lem}[thm]{Lemma}
\newtheorem{prop}[thm]{Proposition}

\newtheorem{rem}{Remark}

\renewcommand\appendix{\par
\setcounter{section}{0}
\setcounter{subsection}{0}
\setcounter{figure}{0}
\setcounter{table}{0}
\renewcommand\thesection{ \Alph{section}}
\renewcommand\thefigure{\Alph{section}\arabic{figure}}
\renewcommand\thetable{\Alph{section}\arabic{table}}
}

\usepackage{babel}

\allowdisplaybreaks

\begin{document}

\title{Wireless Information and Power Transfer over an AWGN channel: Nonlinearity and Asymmetric Gaussian Signalling}

\author{Morteza Varasteh, Borzoo Rassouli, Bruno Clerckx\\
Department of Electrical and Electronic Engineering, Imperial College London, London, U.K.\\
\{m.varasteh12; b.rassouli12;b.clerckx\}@imperial.ac.uk. }

\maketitle

\begin{abstract}
Simultaneous transmission of information and power over a point-to-point flat-fading complex Additive White Gaussian Noise (AWGN) channel is studied. In contrast with the literature that relies on an inaccurate linear model of the energy harvester, an experimentally-validated nonlinear model is considered. A general form of the delivered Direct Current (DC) power in terms of system baseband parameters is derived, which demonstrates the dependency of the delivered DC power on higher order statistics of the channel input distribution. The optimization problem of maximizing Rate-Power (R-P) region is studied. Assuming that the Channel gain is available at both the receiver and the transmitter, and constraining to independent and identically distributed (i.i.d.) channel inputs determined only by their first and second moment statistics, an inner bound for the general problem is obtained. Notably, as a consequence of the harvester nonlinearity, the studied inner bound exhibits a tradeoff between the delivered power and the rate of received information. It is shown that the tradeoff-characterizing input distribution is with mean zero and with asymmetric power allocations to the real and imaginary dimensions.\blfootnote{This work has been partially supported by the EPSRC of the UK, under the grant EP/P003885/1.}
\end{abstract}

\section{Introduction}
Radio-Frequency (RF) waves can be utilized for transmission of both information and power simultaneously. As one of the primary works in the information theory literature, Varshney studied this problem in \cite{Varshney_2008}, in which he characterized the capacity-power function for a point-to-point discrete memoryless channel (DMC). He showed the existence of tradeoff between the information rate and the delivered power for some channels, such as, point-to-point binary channels and amplitude constraint Gaussian channels. Recent results in the literature have also revealed that in many scenarios, there is a tradeoff between information rate and delivered power. Just to name a few, frequency-selective channel \cite{Grover_Sahai_2010}, MIMO broadcasting \cite{Zhang_Keong_2013}, interference channel \cite{Park_Clerckx_2013,Park_Clerckx_2014}, relaying \cite{Nasir_Zhou_Durrani_Kennedy_2013,Huang_Clerckx_2016_2}.

One of the major efforts in a Simultaneous Wireless Information and Power Transfer (SWIPT) architecture is to increase the Direct-Current (DC) power at the output of the harvester without increasing transmit power. The harvester, known as rectenna, is composed of a rectifier\footnote{In the literature, the rectifier is usually considered as a diode, which is the main source of nonlinearity induced in the system.} followed by a low-pass filter. In \cite{Trotter_Griffin_Durgin_2009,Clerckx_Bayguzina_2016}, it is shown that the RF-to-DC conversion efficiency is a function of rectenna's structure, as well as its input waveform. Accordingly, in order to maximize rectenna's DC power output, a systematic waveform design is crucial to make the best use of an available RF spectrum. In \cite{Clerckx_Bayguzina_2016}, an analytical model for rectenna's output is introduced via the Taylor expansion of the diode characteristic function. As one of the main conclusions, it is shown that the rectifier's nonlinearity is key to design efficient wireless powered systems.

The design of an efficient SWIPT architecture fundamentally relies on designing an efficient Wireless Power Transfer (WPT) structure as an important building
block of SWIPT. The SWIPT literature has so far focused on the linear model of the rectifier, e.g., \cite{Grover_Sahai_2010,Zhang_Keong_2013,Park_Clerckx_2013,Park_Clerckx_2014,Nasir_Zhou_Durrani_Kennedy_2013,Huang_Clerckx_2016_2}, whereas, it is expected that considering nonlinearity effect changes the SWIPT design, signalling and architecture significantly. Indeed, in \cite{Clerckx_2016,Clerckx_2016_Proc}, the design of SWIPT waveforms is studied accounting for rectenna's nonlinearity with a power splitter at the receiver. It is shown that superposing deterministic multisines (for power transfer purposes) with Orthogonal Frequency Division Multiplexing (OFDM) symbols modulated with Circularly Symmetric Complex Gaussian (CSCG) zero-mean inputs (for information purposes) enlarges the Rate-Power (R-P) region, compared to merely zero-mean inputs. This highlights the potential and benefits of departing from conventional CSCG inputs in SWIPT.

Leveraging the aforementioned observations, we provide a step closer at identifying the fundamental limits of SWIPT accounting for the nonlinearity of rectenna. In this paper, we study a flat-fading Additive White Gaussian Noise (AWGN) channel for SWIPT. Taking the advantage of the approximation for rectenna's nonlinear output introduced in \cite{Clerckx_Bayguzina_2016}, we obtain the general form of the delivered power in terms of system baseband parameters. Assuming that the receiver jointly extracts information and harvests power from the received RF signal,\footnote{We note that, leveraging the results in thermodynamics of computing, it is demonstrated that energy need not be dissipated in the decoding process. This is due to the reason that to perform a mathematical work, energy is not required \cite[Ch. 5]{Feynman_1998}. In particular, decoders that are reversible computational devices would not dissipate any energy \cite{Landauer_1987} and electronic circuits that are almost thermodynamically reversible have been built \cite{Frank_thesis}. Motivated by this, we also assume that at the receiver, the decoder is able to jointly harvest power and extract information from the received RF signal.} it is shown that the delivered power at the receiver is dependent on the first to fourth moment statistics of the channel input distribution. Considering the optimization problem of maximizing R-P region, we obtain an achievable scheme as an inner bound for the general problem.  The scheme is based on constraining the channel inputs to independent and identically distributed (i.i.d.) distributions that are determined by their first and second moment statistics. For the studied inner bound, we show that there is a tradeoff between the delivered power at the receiver and the rate of the received information. This result is highlighted in contrast to the scenario, in which a linear model is considered for the power harvester at the receiver. It can be easily verified that under an assumption of linear model for the power harvester, the goal of maximum rate and maximum energy are aligned in the flat-fading channel. Additionally, we show that the maximum rate-power (for the studied inner bound) is achieved when the channel input distributions is Gaussian with mean zero, however, with different (asymmetric) power allocations to the real and imaginary dimensions.

\textit{Organization}: In Section \ref{Sec_System_Model}, we introduce the system model. In Section \ref{Sec_Delivered_power}, the delivered power at the receiver is obtained in terms of system baseband parameters accounting the approximation for nonlinearity of rectenna. In Section \ref{Sec_Problem_statement}, we introduce the problem considered in this paper, and accordingly, in Section \ref{Sec_Main_Result}, we obtain an achievable scheme as an inner bound for the general optimization problem. In Section \ref{Sec_Conclusion}, we conclude the paper.

\textit{Notation}: Throughout this paper, the standard CSCG distribution is denoted by $\mathcal{CN}(0,1)$. Complex conjugate of a complex number $c$ is denoted by $c^{*}$. For a random process $X(t)$, corresponding random variable at time index $n$ is represented by $X_n$. The operators $\mathbb{E}[\cdot]$ and $\mathcal{E}[\cdot]$ denote the expectation over statistical randomness and the average over time, respectively. $\Re\{\cdot\}$ and $\Im\{\cdot\}$ are real and imaginary operators, respectively. We use the notations $\mathrm{sinc}(t)=\frac{\sin(\pi t)}{\pi t}$ and $s_l=\text{sinc}(l+1/2)$ for integer $l$. We also define $\delta_l$ as
\begin{align}
\delta_l=\left\{\begin{array}{ll}
  1 & l=0 \\
  0 & l\neq 0
\end{array}\right..
\end{align}

\section{System Model}\label{Sec_System_Model}
Considering a point-to-point flat-fading AWGN channel, in the following, we explain the operation of the transmitter and the receiver.

\subsection{Transmitter}
At the transmitter, the signal $X(t)$ is produced as
\begin{align}
X(t)=\sum_{n}X_n\text{sinc}(f_wt-n),
\end{align}
where $X_n$ is an information-power symbol at time index $n$, modelled as a random variable, which is produced in an i.i.d. fashion and $X(t)$ is with bandwidth $[-f_w/2,f_w/2]$. Next, the signal $X(t)$ is upconverted to the carrier frequency $f_c$ and is sent over the channel.

\subsection{Receiver}
The filtered received RF waveform at the receiver is modelled as
\begin{align}
  Y_{\text{rf}}(t) &=\sqrt{2}\Re\left\{Y(t)e^{j2\pi f_ct}\right\},
\end{align}
where $Y(t)$ is the baseband equivalent of the channel output with bandwidth $[-f_w/2,f_w/2]$. We assume that $f_c>2f_w$.

\textit{Power}: At the receiver, the power of the RF signal $Y_{\text{rf}}(t) $ is delivered through the rectenna. In the following, we leverage the approximation for rectenna's output introduced in \cite{Clerckx_Bayguzina_2016}\footnote{According to \cite{Clerckx_Bayguzina_2016}, due to the presence of a diode in rectenna's structure, its output current is an exponential function, which is approximated by expanding its Taylor series. The approximation used here, is the fourth moment truncation of Taylor series, in which the first and third moments are zero with respect to the time averaging.}. Accordingly, the delivered power (denoted by $P_{\text{del}}$) is modelled as\footnote{According to \cite{Clerckx_Bayguzina_2016}, rectenna's output in (\ref{eqn:1}) is in the form of current with unit Ampere. However, since power is proportional to current, with abuse of notation, we refer to the term in (\ref{eqn:1}) as power.}
\begin{align}\label{eqn:1}
P_{\text{del}}=\mathbb{E}\mathcal{E}[k_2Y_{\text{rf}}(t)^2 + k_4 Y_{\text{rf}}(t)^4],
\end{align}
where $k_2$ and $k_4$ are constants. Note that, in the linear model for the delivered power $P_{\text{del}}$, in (\ref{eqn:1}), we have only the second moment of the received RF signal $Y_{\text{rf}}(t)$. Validating through circuit simulations in \cite{Clerckx_Bayguzina_2016}, it is shown that the linear model is inaccurate and inefficient from a signal design perspective.

\textit{Information}: The signal $Y_{\text{rf}}(t) $ is downconverted producing the baseband signal $Y(t)$ given as \footnote{We model the baseband equivalent channel impulse response as $H(\tau,t)=\sum_{i}a_i^b(t)\delta(\tau-\tau_i(t))+W(t)$ where $\alpha_i^b(t)$, $\tau_i(t)$ are the channel coefficient and delay of path $i$.}
\begin{align}
Y(t)=\sum_{i}a_i^b(t)X(t-\tau_i(t))+W(t).
\end{align}
Next, $Y(t)$ is sampled with a sampling frequency $f_w$ producing $Y_{m}=Y(m/f_w)$ given as
\begin{align}\label{eqn:20}
Y_m&=X_m \sum_{i}a_i^b(m/f_w)+W_m,
\end{align}
where in (\ref{eqn:20}), we used $\tau_i(m/f_w)f_w\approx 0$ because the channel is flat-fading. $W_m$ and $X_m$ represent samples of the additive noise $W(t)$ and the signal $X(t)$ at time $t=m/f_w$, respectively.

We model $W_m$ as an i.i.d. and CSCG random variable with variance $\sigma_w^2$, i.e., $W_m\sim \mathcal{CN}(0,\sigma_w^2)$. We assume that both the transmitter and the receiver know the Channel gain, namely, $h(t)=\sum_{i}a_i^b(t)$ at times $t=m/2f_w$ for integer $m$, which is assumed to be fixed over all the transmissions. Throughout the paper, since the transmitted symbols $X_m$ and the noise $W_m$ are i.i.d., we drop the index $m$ for $X_m,~W_m$. We also define $h=\sum_i a_i^b((2m)/(2f_w))$ and $\tilde{h}=\sum_i a_i^b((2m+1)/(2f_w))$. Note that $h$ and $\tilde{h}$ are assumed to be fixed, however, we assume they are not necessarily equal. Therefore, (\ref{eqn:20}) reads
\begin{align}\label{eqn:34}
Y=h X + W.
\end{align}
Note that in (\ref{eqn:34}), only even samples of the channel, i.e., $h$ are involved.

\section{Delivered power}\label{Sec_Delivered_power}
In this section, we study the power delivered at the receiver. Note that most of the communication processes, such as, coding/ decoding, modulation/ demodulation, etc, is done at the baseband. Therefore, from a communication system design point of view, it is most preferable to have baseband equivalent presentation of the system. Henceforth, in the following Proposition, we derive the delivered power $P_{\text{del}}$ at the receiver in terms of system baseband parameters.

\begin{prop}\label{Prop1}
Assuming the channel input distributions are i.i.d., the delivered power $P_{\text{del}}$ at the receiver, can be expressed in terms of system baseband parameters as
\begin{align}\label{eqn:22}
P_{\text{del}}=\alpha Q+\tilde{\alpha}\tilde{Q}+(\beta +\tilde{\beta})P+\gamma,
\end{align}
where $\tilde{Q}$ is given by
\begin{align}\nonumber
\tilde{Q}&=\frac{1}{3}\big(Q_{r}+Q_{i}+2(\mu_{r}T_{r}+\mu_{i}T_{i})\\
&+6P_{r}P_{i}+6P_{r}(P_{r}-\mu_{r}^{2})+6P_{i}(P_{i}-\mu_{i}^{2})\big),
\end{align}
where the parameters $\alpha,~\tilde{\alpha},~\beta,~\tilde{\beta}$ and $\gamma$ are given as
\begin{align}\label{eqn:32}
\alpha&=\frac{3k_4}{4f_w}|h|^4,\\
\tilde{\alpha}&=\frac{3k_4}{4f_w}|\tilde{h}|^4,\\
 \beta&=\frac{1}{f_w}\left(k_2+6k_4\sigma_w^2\right)|h|^2, \\
 \tilde{\beta}&=\frac{1}{f_w}\left(k_2+6k_4\sigma_w^2\right)|\tilde{h}|^2, \\\label{eqn:33}
 \gamma&=\frac{1}{f_w}(k_2\sigma_w^2+3k_4\sigma_w^4),
\end{align}
and $Q=\mathbb{E}[|X|^4]$, $T=\mathbb{E}[|X|^3]$, $P=\mathbb{E}[|X|^2]$, $\mu=\mathbb{E}[X]$. Similarly, $Q_r=\mathbb{E}[\Re\{X\}^4]$, $T_r=\mathbb{E}[\Re\{X\}^3]$, $P_r=\mathbb{E}[\Re\{X\}^2]$, $\mu_r=\mathbb{E}[\Re\{X\}]$ and $Q_i=\mathbb{E}[\Im\{X\}^4]$, $T_i=\mathbb{E}[\Im\{X\}^3]$, $P_i=\mathbb{E}[\Im\{X\}^2]$, $\mu_i=\mathbb{E}[\Im\{X\}]$.
\end{prop}
\textit{Proof}: See Appendix \ref{app:1}.

\begin{rem}
We note that obtaining a closed form expression for the delivered power $P_{\text{del}}$ at the receiver when the channel inputs are not i.i.d. is cumbersome. This is due to the fact that the fourth moment of the received signal $Y_{\text{rf}}(t)$ creates dependencies of the statistics of the present channel input on the statistics of the channel inputs on other time indices (see e.g., eq. (\ref{eqn:9}) and eq. (\ref{eqn:30}) in Appendix \ref{app:1}).
\end{rem}

\section{Problem statement}\label{Sec_Problem_statement}

We aim at maximizing the rate of the received information, as well as the amount of power delivered at the receiver. Accordingly, the optimization problem we consider, is the maximization of mutual information between the channel input $X$ and the channel output $Y$ under a given power constraint at the transmitter and a minimum delivered power constraint at the receiver. Hence, for the optimization problem, we have
\begin{equation}\label{eqn:13}
\begin{aligned}
& \underset{ p_{X}(x)}{\text{sup}}
& & I\left(X;Y\right) \\
& \text{s.t.}
& & \left\{\begin{array}{l}
      P\leq P_{a} \\
      P_{\text{del}}\geq P_d
    \end{array}\right.,
\end{aligned}
\end{equation}
where $\sup$ is taken over all input distributions $p_{X}(x)$ satisfying the constraints in (\ref{eqn:13}). $P_{a}$ is the available power budget at the transmitter and $P_d$ is the minimum amount of power that is to be delivered to the receiver.

\begin{rem}
We note that, for the problem in (\ref{eqn:13}), if the second constraint (the minimum delivered power at the receiver) is represented via a linear model, i.e., $\mathbb{E}[|Y|^2]\geq P_d$, the maximum is achieved using a CSCG input distribution. It can also be verified easily that there is no tradeoff between the received information rate and delivered power at the receiver.
\end{rem}

\section{Main Result}\label{Sec_Main_Result}
In this section, we obtain an inner bound for the problem in (\ref{eqn:13}) by constraining the input distributions to those that are determined by their first and second moment statistics\footnote{This assumption is justified due to the fact that in practice, most of the modulation schemes are i.i.d. and are fully characterized by the knowledge of the first and second moment statistics only.}. We show that for the considered scenario, there is a tradeoff between the rate of the transmitted information, namely $I(X;Y)$ and delivered power $P_{\text{dc}}$ at the receiver and accordingly, we characterize the tradeoff.

\begin{prop}\label{Prop2}
When a channel input distribution $p_{X}(x)$ is completely determined by its first and second moment statistics, the supremum in (\ref{eqn:13}) is achieved by a zero mean Gaussian distribution as the channel input, i.e., $\Re\{X\}\sim\mathcal{N} (0,P_r)$, and $\Im\{X\}\sim \mathcal{N}(0,P_i)$, where $P=P_r+P_i=P_a$. Furthermore, let $P_{\text{dc,max}}=3(\alpha+\tilde{\alpha}) {P_{a}}^2+(\beta+\tilde{\beta}) P_{a} +\gamma$ and $P_{\text{dc,min}}=2(\alpha+\tilde{\alpha}) {P_{a}}^2+(\beta+\tilde{\beta}) P_{a} +\gamma$ be the maximum and minimum delivered power at the receiver, respectively. For $P_d=P_{\text{dc,max}}$, the maximum in (\ref{eqn:13}) is attained by $P_i=0,P_r=P_{a}$ or $P_i=P_a,P_r=0$. For $P_d=P_{\text{dc,min}}$, the maximum in (\ref{eqn:13}) is attained by $P_i=P_a/2,P_r=P_a/2$. For $P_{\text{dc,min}}<P_d< P_{\text{dc,max}}$, the optimal power allocation that attains the maximum rate is given by $P_i^*$ and $P_r^*=P_a-P_i^*$, where $P_i^*$ is chosen such that the following equation is satisfied
\begin{align}\label{eqn:31}
\alpha Q+\tilde{\alpha} \tilde{Q}+(\beta+\tilde{\beta}) P_a+\gamma=P_d.
\end{align}

For $P_d< P_{\text{dc,min}}$, the optimal power allocation is attained by $P_i^*=P_r^*=P_{a}/2$ and the delivered power is still $P_{\text{dc,min}}$.

\end{prop}
\textit{Proof}: See Appendix \ref{app:2}.

Note that in (\ref{eqn:31}), for a complex zero-mean Gaussian distributed channel input with $P_r$ and $P_i$ as the variances of real and imaginary dimensions, respectively, we have
\begin{align}
Q=\tilde{Q}=3(P_i^2+P_r^2)+2P_iP_r.
\end{align}

\begin{figure}
\begin{centering}
\includegraphics[scale=0.3]{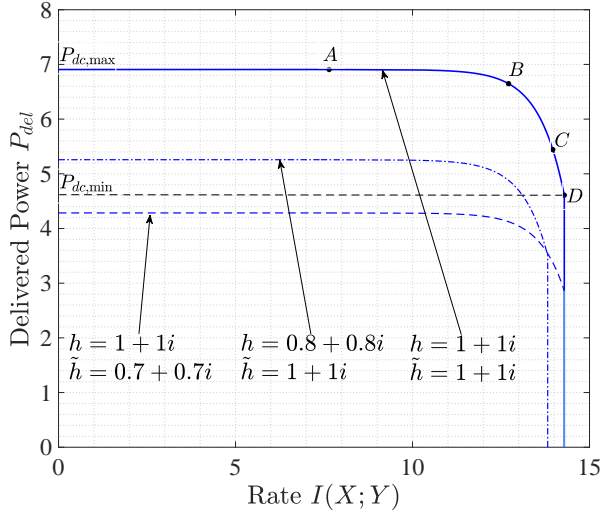}
\caption{Delivered power $P_{\text{del}}$ and achievable information rate $I(X;Y)$ for the problem in (\ref{eqn:13}) for i.i.d. channel inputs that are determined by their first and second moment statistics $(P_{a}=1,~ \sigma_w^2=10^{-4},~f_w=1,~k_2=0.17,~k_4=19.145$). The points $A,~B,~C,~D$ correspond to $(P_r,P_i)=(0,1),~=(0.03,0.97),~=(0.2,0.8),~=(0.5,0.5)$, respectively, where the values for $P_r$ and $P_i$ can be interchanged without affecting the resulting points. The values for $k_2,~k_4$ are adapted from \cite{Clerckx_2016}.} \label{Figure:1}
\par\end{centering}
\vspace{0mm}
\end{figure}

\begin{rem}
From (\ref{eqn:22}), it is seen that the delivered power $P_{\text{del}}$ at the receiver depends on the second moment statistics $P_r,P_i$, as well as the fourth moment statistics $Q_r,Q_i$ of the channel input $X$. This is due to the presence of the fourth moment of the received signal $Y_{\text{rf}}$ in modelling the rectenna's output. Accordingly, the rate is minimized (corresponding to $P_{\text{dc,max}}$) when the available power at the transmitter is fully allocated to one of the real or imaginary dimensions. This is because allocating power to one dimension, leads to a higher fourth order moment. On the other hand, the maximum rate is achieved (corresponding to $P_{\text{dc,min}}$) when the available power is equally distributed between the real and the imaginary dimensions. This is elaborated in Figure  \ref{Figure:1}.
\end{rem}

\begin{rem}
As mentioned earlier, maximization of both the delivered power $P_{\text{del}}$ and the rate $I(X;Y)$ are aligned under the linear modelling for the delivered power. Therefore, regardless of the channel condition, the best power allocation is always $P_i=P_r=P/2$. However, accounting the fourth moment of the received signal $Y_{\text{rf}}$ in (\ref{eqn:1}), the receiver chooses the proper power allocation such that the constraints in (\ref{eqn:13}) are satisfied (if $P_d\leq P_{\text{dc,max}}$). We also note that since we have assumed the channel is fixed for the whole transmission, therefore, the transmitter keeps using the same power allocation for the whole transmission.

It is also noted that at each time index the rate of the information $I(X;Y)$ is affected only through $h$, whereas, the delivered power $P_{\text{del}}$ is affected through both $h$ and $\tilde{h}$. This is illustrated in Figure \ref{Figure:1} by representing three different regions by varying the channel coefficients.
\end{rem}

\section{Conclusion}\label{Sec_Conclusion}
In this paper, we studied SWIPT over a point-to-point complex AWGN channel in the presence of a nonlinear power harvester at the receiver. Assuming that the channel state information is available at both the transmitter and the receiver, we studied the problem of maximizing rate of the transmitted information as well as delivered power at the receiver. Assuming that the channel inputs are i.i.d. and are fully characterized by the knowledge of their first and second moment statistics, we derived an inner bound for the optimal R-P region. We showed that for the obtained inner bound, there is a trade off (due to the nonlinearity of the power harvester at the receiver) between the rate of transmitted information and delivered power. Accordingly, we characterized the inner bound, which demonstrates that the optimal channel input is still a zero mean Gaussian distribution, however, with asymmetric power allocations to the real and imaginary dimensions.

Among open problems that are left for future research, we mention here the optimal input distribution for the problem in (\ref{eqn:13}). Another interesting problem is the extension of the problem studied in this paper to the frequency-selective AWGN channel.
\appendix
\section{Proof of Proposition \ref{Prop1}} \label{app:1}
The following series will be useful throughout the proof of the proposition \ref{Prop1}.
\begin{lem}\label{Lemma1} Recalling that $s_l=\text{sinc}(l+1/2)$ for integer $l$, we have the following series:
\begin{align}
S_0&\triangleq \sum_{l} s_l^2=1,\\
S_1&\triangleq \sum_{l}\sum_{k:k\neq l} s_l  s_k=0,\\
S_2&\triangleq \sum_{l}\sum_{k:k\neq l}\sum_{\substack{d:d\neq l\\d\neq k}}\sum_{\substack{m:m\neq l\\m\neq d\\m\neq k}} s_l  s_k s_d s_m=0,\\
S_3&\triangleq\sum_{l}\sum_{k:k\neq l} s_l^2  s_k^2=\frac{2}{3},\\
S_4&\triangleq\sum_{l}\sum_{k:k\neq l}\sum_{\substack{d:d\neq l\\d\neq k}} s_l^2  s_k s_d=-\frac{1}{3},\\
S_5&\triangleq\sum_{l} s_l^4=\frac{1}{3},\\
S_6&\triangleq\sum_{l}\sum_{k:k\neq l}s_l^3  s_k=\frac{1}{2}.
\end{align}

\end{lem}
\textit{Proof}: See Appendix \ref{app:3}.

Considering first the term $\mathbb{E}\mathcal{E}[Y_{\text{rf}}(t)^2]$, we have
\begin{align}
  \mathbb{E}\mathcal{E}&[Y_{\text{rf}}(t)^2] =\frac{1}{2} \mathbb{E}\mathcal{E}\left[\left(Y(t)e^{jf_ct}+Y^{*}(t)e^{-jf_ct}\right)^2\right] \\\label{eqn:2}
  &=\mathbb{E}\mathcal{E}\left[|Y(t)|^2\right]\\\label{eqn:3}
  &=\mathbb{E}\mathcal{E}\bigg[\sum\limits_{n,m}Y_nY_m^* \mathrm{sinc}(f_wt-n)\mathrm{sinc}(f_wt-m)\bigg]\\
  &=\sum\limits_{n,m}\mathbb{E}\left[Y_nY_m^*\right] \mathcal{E}\left[\mathrm{sinc}(f_wt-n)\mathrm{sinc}(f_wt-m)\right]\\\label{eqn:4}
 &=\lim_{T\rightarrow\infty}\frac{1}{f_wT}\sum\limits_{m}\mathbb{E}\left[|Y_m|^2\right]\\\label{eqn:29}
 &=|h|^2P+\sigma_w^2,
\end{align}
where (\ref{eqn:2}) is because we have $\mathcal{E}\{Y(t)^2e^{2jf_ct}\}=0$. (\ref{eqn:3}) is due to the fact that the signal $Y(t)$ is bandlimited to $f_w$ and we have
\begin{align}
Y(t)=\sum_{n}Y_n\text{sinc}(f_wt-n).
\end{align}
In (\ref{eqn:4}), we used the equation
\begin{align}
\mathcal{E}\left[\mathrm{sinc}(f_wt-n)\mathrm{sinc}(f_wt-m)\right]=\lim_{T\rightarrow\infty}\frac{1}{f_wT}\delta_{n-m}.
\end{align}

Considering the term $\mathbb{E}\mathcal{E}[Y_{\text{rf}}(t)^4]$, we have
\begin{align}\nonumber
  \mathbb{E}\mathcal{E}[Y_{\text{rf}}(t)^4] &=\frac{1}{4} \mathbb{E}\mathcal{E}\big[4|Y(t)|^4\\\nonumber
  &+(Y(t)^2e^{j2f_ct}+Y^{*}(t)^2e^{-j2f_ct})^2 \\
  &+4|Y(t)|^2(Y(t)^2e^{j2f_ct}+Y^{*}(t)^2e^{-j2f_ct})\big] \\\label{eqn:5}
  &=\frac{3}{2}\mathbb{E}\mathcal{E}\left[|Y(t)|^4\right].
\end{align}
Note that, the signal $|Y(t)|^2$ is real with bandwidth $[-f_w,f_w]$. Hence, it can be represented by its samples taken each $t=1/2f_w$ seconds. Therefore, we have
\begin{align}
|Y(t)|^2=\sum_{n}Y^s_n\text{sinc}(2f_wt-n),
\end{align}
where $Y^s_n\triangleq |Y(n/2f_w)|^2$. Accordingly, (\ref{eqn:5}) reads as
\begin{align}
\frac{3}{2}\mathbb{E}\mathcal{E}&\left[|Y(t)|^4\right]=\frac{3}{2f_w}\sum_{n}\mathbb{E}[|Y^s_n|^2]\\\label{eqn:6}
&=\lim_{T\rightarrow\infty}~\frac{3}{2Tf_w}\sum_{k}\mathbb{E}[|Y^s_{2k+1}|^2]+\frac{3}{2Tf_w}\sum_{k}\mathbb{E}[|Y^s_{2k}|^2].
\end{align}
Note that $Y^s_{2k}= |Y(2k/2f_w)|^2= |Y_k|^2$. Hence, $\mathbb{E}[|Y^s_{2k}|^2]$ in (\ref{eqn:6}) reads
\begin{align}
&\mathbb{E}[|Y^s_{2k}|^2]=\mathbb{E}[|Y_k|^4]\\
&=\mathbb{E}[((hX+W)(h^*X^*+W^*))^2]\\\nonumber
&=\mathbb{E}[|h|^4|X|^4+|W|^4+2|h|^2|X|^2|W|^2\\\nonumber
&\quad+W^2h^{*2}X^{*2}+W^{*2}h^2X^2+2|h|^2|X|^2|W|^2\\\nonumber
&\quad+2(|h|^2|X|^2Wh^*X^*+|h|^2|X|^2W^*hX\\
&\quad+|W|^2Wh^*X^*+|W|^2W^*hX)]\\\nonumber
&=|h|^4\mathbb{E}[|X|^4]+2\sigma_w^4\\
&\quad+2\sigma_w^2|h|^2\mathbb{E}[|X|^2]+2\sigma_w^2|h|^2\mathbb{E}[|X|^2]\\\label{eqn:7}
&=|h|^4Q+4\sigma_w^2|h|^2P+2\sigma_w^4.
\end{align}

To calculate the term $\mathbb{E}[|Y^s_{2k+1}|^2]$ in (\ref{eqn:6}), we note that the channel's baseband equivalent signal $Y(t)$ can be written as
\begin{align}
Y(t)=\sum_{n}X_n\sum_{i}a_i^b(t)\text{sinc}(f_wt-n)+W(t),
\end{align}
where we have neglected the term $f_w \tau_i$, since the channel is flat and we have $f_w \tau_i\approx 0$. Substituting $t=(2k+1)/f_w$ we have
\begin{align}
\tilde{Y}_k&\triangleq Y\left(\frac{2k+1}{2f_w}\right)\\
&=\sum_{n}X_n\sum_{i}a_i^b\left(\frac{2k+1}{2f_w}\right)s_{k-n}+W\left(\frac{2k+1}{2f_w}\right)\\
&=\sum_{n}X_ns_{k-n}\sum_{i}a_i^b\left(\frac{2k+1}{2f_w}\right)+W\left(\frac{2k+1}{2f_w}\right)\\
&=\tilde{h}\tilde{X}+\tilde{W}.
\end{align}
where $\tilde{X}\triangleq \sum_{n}X_ns_{k-n}$ and $\tilde{W}\triangleq W((2k+1)/2f_w)$. Similarly to (\ref{eqn:7}), we have
\begin{align}
\mathbb{E}[|Y^s_{2k+1}|^2]&=\mathbb{E}[|\tilde{Y}_k|^4]\\\label{eqn:8}
&=|\tilde{h}|^4\tilde{Q}+4\sigma_w^2|\tilde{h}|^2\tilde{P}+2\sigma_w^4,
\end{align}
where $\tilde{Q}=\mathbb{E}[|\tilde{X}|^4],~\tilde{P}=\mathbb{E}[|\tilde{X}|^2]$. For $\tilde{P}$, we have
\begin{align}\label{eqn:30}
\tilde{P}&=\mathbb{E}\bigg[\sum_{n,m}X_nX_m^*s_{k-n}s_{k-m}\bigg]\\\nonumber
&=\sum_{n,m:n=m}\mathbb{E}[|X_n|^2]s_{k-n}^2\\\label{eqn:26}
&\quad+\sum_{n,m:n\neq m}\mathbb{E}[X_n]\mathbb{E}[X_m^*]s_{k-n}s_{k-m}\\
&=S_0P+S_1|\mu|^2\\\label{eqn:28}
&=P,
\end{align}
where in (\ref{eqn:26}) we used the assumption that $X_n$ is i.i.d. with respect to different values of $n$.
For $\tilde{Q}$, we have
\begin{align}\label{eqn:9}
\tilde{Q}&=\mathbb{E}\bigg[\sum_{l,k,d,m}X_lX^*X_d^*X_m^*s_{n-l}s_{n-k}s_{n-d}s_{n-m}\bigg].
\end{align}

Accounting for the different cases for the possible values of $l,k,d,m$, we have
\begin{itemize}
\item If all the indices $l,k,d,m$ are with different values, we have
\begin{align}
\tilde{Q}=|\mu|^4S_2.
\end{align}
\item If $(l=k,~d\neq k,~d=m)$ or $(l=d,~k\neq d,~k=m)$, we have
\begin{align}
\tilde{Q}=P^2S_3.
\end{align}
\item If $(l=m,~k\neq m,~k=d)$, we have
\begin{align}
\tilde{Q}=|\bar{P}|^2S_3.
\end{align}
\item If $(l=k,~d\neq m,~d\neq k,~m\neq k)$ or $(l=d,~k\neq m,~k\neq d,~m\neq d)$ or $(k=m,~l\neq d,~l\neq m,~d\neq m)$ or $(d=m,~l\neq k,~l\neq m,~k\neq m)$, we have
\begin{align}
\tilde{Q}=P|\mu|^2S_4.
\end{align}
\item If $(l=m,~k\neq d,~k\neq m,~d\neq m)$, we have
\begin{align}
\tilde{Q}=\bar{P}\mu^{*2}S_4.
\end{align}
\item If $(k=d,~l\neq m,~l\neq d,~m\neq d)$, we have
\begin{align}
\tilde{Q}=\bar{P}^*\mu^{2}S_4.
\end{align}
\item If $l=k=d=m$, we have
\begin{align}
\tilde{Q}=QS_5.
\end{align}
\item If $l=k=d\neq m$ or $k=d=m\neq l$, we have
\begin{align}
\tilde{Q}=\bar{T}^*\mu S_6.
\end{align}
\item If $l=d=m\neq k$ or $l=k=m\neq d$, we have
\begin{align}
\tilde{Q}=\bar{T}\mu^*S_6.
\end{align}
\end{itemize}
In the above expressions we define $\bar{P}=\mathbb{E}[X^2]$, $\bar{T}=\mathbb{E}[|X|^2X]$. Hence, (\ref{eqn:9}) reads
\begin{align}\nonumber
\tilde{Q}&=|\mu|^4S_2+(2P^2+|\bar{P}|^2)S_3\\\nonumber
&\quad+(4P|\mu|^2+\bar{P}\mu^{*2}+\bar{P}^*\mu^2)S_4\\
&\quad+QS_5+2(\bar{T}\mu^*+\bar{T}^*\mu)S_6\\\nonumber
&=\frac{1}{3}\bigg[Q+4P(P-|\mu|^2)\\\label{eqn:10}
&\quad+2(|\bar{P}|^2-\Re\{\bar{P}\mu^{*2}\})+2\Re\{\bar{T}\mu^*\}\bigg].
\end{align}

Expanding the terms $|\bar{P}|^2-\Re\{\bar{P}\mu^{*2}\}$ and $\Re\{\bar{T}\mu^*\}$ in (\ref{eqn:10}), we have
\begin{align}\label{eqn:11}
|\bar{P}|^2-\Re\{\bar{P}\mu^{*2}\}&=(P_r-P_i)(P_r-P_i-(\mu_r^2-\mu_i^2)),\\\label{eqn:12}
\Re\{\bar{T}\mu^*\}&=\mu_r(T_r+\mu_rP_i)+\mu_i(T_i+\mu_iP_r).
\end{align}

Noting that $Q=Q_i+Q_r+2P_rP_i$ and substituting in (\ref{eqn:10}) along with (\ref{eqn:11}) and (\ref{eqn:12}), after some manipulations $\tilde{Q}$ reads
\begin{align}\nonumber
\tilde{Q}&=\frac{1}{3}\big(Q_r+Q_i+2(\mu_rT_r+\mu_iT_i)\\\label{eqn:27}
&+6(P_rP_i+P_r(P_r-\mu_r^2)+P_i(P_i-\mu_i)\big).
\end{align}

Substituting (\ref{eqn:27}), (\ref{eqn:28}) in (\ref{eqn:8}) and substituting the result along with (\ref{eqn:7}) in (\ref{eqn:6}), and adding with (\ref{eqn:29}) yields the result of the Proposition.

\section{Proof of Proposition \ref{Prop1}} \label{app:2}
Note that constraining the input distributions $p_{X}(x)$ to those that are determined by their first and second moment statistics, the supremum in (\ref{eqn:13}) is attained in general by a non-zero mean Gaussian distribution for each dimension, i.e., $\Re\{X\}\sim (\mu_r,\sigma_r^2)$ and $\Im\{X\}\sim (\mu_i,\sigma_i^2)$, where $\sigma_r^2\triangleq P_r-\mu_r^2$ and $\sigma_i^2\triangleq P_i-\mu_i^2$. Therefore, the optimization problem in (\ref{eqn:13}) reads
\begin{equation}\label{eqn:14}
\begin{aligned}
& \underset{ \mu_r,\mu_i,P_r,P_i}{\text{max}}
& & \frac{f_w}{2}\left(\log(1+a\sigma_r^2)+\log(1+a\sigma_i^2)\right) \\
& \text{s.t.}
& &\!\!\!\!\!\!\!\!\! \left\{\begin{array}{l}
      P_r+P_i\leq P_{a}\\
      \alpha Q+\tilde{\alpha} \tilde{Q}+(\beta+\tilde{\beta}) P+\gamma\geq P_d\\
      \sigma_r^2\geq 0 ,~\sigma_i^2\geq 0
    \end{array}\right.,
\end{aligned}
\end{equation}
where $a\triangleq 2|h|^2/f_w\sigma_w^2$. Writing the K.K.T. conditions for the optimization problem in (\ref{eqn:14}), we have
\begin{align}\label{eqn:17}
      &\lambda_1(P_r+P_i-P_{a})=0,~ \lambda_1\geq0 \\
      &\lambda_2(\alpha Q+\tilde{\alpha} \tilde{Q}+(\beta+\tilde{\beta}) P+\gamma-P_d)=0,~ \lambda_2\geq0, \\
      &\zeta_r\sigma_r^2=0,~\zeta_i\sigma_i^2=0,~\zeta_r,\zeta_i\geq 0\\\nonumber
      &\zeta_r=\frac{-c_1a}{1+a\sigma_r^2}+\lambda_1\\\label{eqn:16}
      &\quad-\lambda_2 (2(\alpha+\tilde{\alpha})(3P_r+P_i)+\beta+\tilde{\beta}),\\\nonumber
      &\zeta_i=\frac{-c_1a}{1+a\sigma_i^2}+\lambda_1\\\label{eqn:21}
      &\quad-\lambda_2 ((\alpha+\tilde{\alpha})(3P_i+P_r)+\beta+\tilde{\beta}),\\\label{eqn:15}
      &\frac{2c_1a\mu_r}{1+a\sigma_r^2}+8\lambda_2 (\alpha+\tilde{\alpha})\mu_r^3+2\zeta_r\mu_r=0,\\\label{eqn:18}
      &\frac{2c_1a\mu_i}{1+a\sigma_i^2}+8\lambda_2 (\alpha+\tilde{\alpha}) \mu_i^3+2\zeta_i\mu_i=0,
\end{align}
where $c_1\triangleq(f_w \log e)/2$ and in (\ref{eqn:16}) to (\ref{eqn:18}) we used the following
\begin{align}
&\frac{\partial Q}{\partial P_r}=\frac{\partial \tilde{Q}}{\partial P_r}=6P_l+2P_i,\\
&\frac{\partial Q}{\partial P_i}=\frac{\partial \tilde{Q}}{\partial P_i}=6P_i+2P_l,\\
&\frac{\partial Q}{\partial \mu_r}=\frac{\partial \tilde{Q}}{\partial \mu_r}=-8\mu_r^3,\\
&\frac{\partial Q}{\partial \mu_i}=\frac{\partial \tilde{Q}}{\partial \mu_i}=-8\mu_i^3.
\end{align}

It can be easily verified from (\ref{eqn:17}), (\ref{eqn:16}) and (\ref{eqn:21}) that when $\lambda_2=0$, the maximum is achieved when $\mu_r=\mu_i=0$ and $P_r=P_i=\frac{P_{a}}{2}$, yielding $P_{\text{dc,min}}=2(\alpha+\tilde{\alpha}) {P_{a}}^2+(\beta+\tilde{\beta}) P_{a} +\gamma$. For positive values of $\lambda_2$ from (\ref{eqn:16}) it is verified that $\lambda_1>0$, which from (\ref{eqn:17}) results that $P_r+P_i=P_{a}$. The condition $P_r+P_i=P_{a}$ reduces the number of variables $P_i,P_r$ to one. Accordingly, since the mutual information $I(X;Y)$ is concave w.r.t. $P_i\in[0,P_a]$ attaining its maximum at $P_i=P_a/2$ and the delivered power $P_{\text{del}}$ is convex w.r.t. $P_i\in[0,P_a]$ attaining its maximum at $P_i=0$ or $P_i=P_a$ the Proposition is proved.

\section{Proof of Proposition \ref{Lemma1}} \label{app:3}
To prove the series, we will use the following special cases of  Riemann's zeta function and alternating series \cite[Sec. 9.5]{Gradshteyn:book}
\begin{align}
&\sum_{l=1}^{\infty}\frac{1}{l^2}=\frac{\pi^2}{6},\\
&\sum_{l=1}^{\infty}\frac{1}{l^4}=\frac{\pi^4}{90},\\
&\sum_{l=0}^{\infty}\frac{(-1)^l}{(2l+1)^3}=\frac{\pi^3}{32},\\
&\sum_{l=0}^{\infty}\frac{(-1)^l}{(2l+1)}=\frac{\pi}{4}.
\end{align}

We have
\begin{align}
T_0&=\sum_l s_l\\
&=\frac{1}{\pi}\sum_{l}\frac{(-1)^l}{\left(\frac{1}{2}+l\right)}\\
&= \frac{2}{\pi}\bigg[\sum_{l=-\infty}^{-1}\frac{(-1)^l}{(2l+1)}+\sum_{0}^{\infty}\frac{(-1)^l}{(2l+1)}\bigg]\\\label{eqn:19}
&=\frac{2}{\pi}\bigg(\frac{\pi}{4}+\frac{\pi}{4}\bigg)=1,\\
S_0&=\sum_l s_l^2\\
&=\sum_{l}\frac{(-1)^{2l}}{\pi^2 \left(\frac{1}{2}+l\right)^2}\\
&=\frac{4}{\pi^2}\sum_l\frac{1}{(2l+1)^2}\\
&= \frac{4}{\pi^2}\bigg[\sum_{l=-\infty}^{-1}\frac{1}{(2l+1)^2}+\sum_{0}^{\infty}\frac{1}{(2l+1)^2}\bigg]\\
&=\frac{4}{\pi^2}\bigg(\frac{\pi^2}{8}+\frac{\pi^2}{8}\bigg)=1,\\
T_1&=\sum_l s_l^3\\
&=\sum_{l}\frac{(-1)^{3l}}{\pi^3 \left(\frac{1}{2}+l\right)^3}\\
&=\frac{8}{\pi^3}\bigg[\sum_{l=-\infty}^{-1}\frac{1}{(2l+1)^3}+\sum_{0}^{\infty}\frac{1}{(2l+1)^3}\bigg]\\
&=\frac{8}{\pi^3}\bigg(\frac{\pi^3}{32}+\frac{\pi^3}{32}\bigg)=\frac{1}{2},\\
S_5&=\sum_l s_l^4\\
&=\sum_{l}\frac{(-1)^{4l}}{\pi^4 \left(\frac{1}{2}+l\right)^4}\\
&=\frac{16}{\pi^4}\sum_l\frac{1}{(2l+1)^4}\\
&= \frac{16}{\pi^4}\bigg[\sum_{l=-\infty}^{-1}\frac{1}{(2l+1)^4}+\sum_{0}^{\infty}\frac{1}{(2l+1)^4}\bigg]\\
&=\frac{16}{\pi^4}\bigg(\frac{\pi^4}{96}+\frac{\pi^4}{96}\bigg)=\frac{1}{3},\\
S_1&=\sum_{l} \sum_{k,k\neq l}s_ls_k\\
&=\sum_{l} s_l\bigg(\sum_{k}s_k-s_l\bigg)\\
&=\left(\sum_l s_l\right)^2-\sum_l s_l^2\\
&=1-1=0,\\
S_3&=\sum_{l} \sum_{k,k\neq l}s_l^2s_k^2\\
&=\sum_{l} s_l^2\bigg(\sum_{k}s_k^2-s_l^2\bigg)\\
&=\left(\sum_l s_l^2\right)^2-\sum_l s_l^4\\
&=1-\frac{1}{3}=\frac{2}{3},\\
S_6&=\sum_{l} \sum_{k,k\neq l}s_l^3s_k\\
&=\sum_{l} s_l^3\bigg(\sum_{k}s_k-s_l\bigg)\\
&=\frac{1}{2}-\frac{1}{3}=\frac{1}{6},\\
S_4&=\sum_{l}\sum_{k,k\neq l}\sum_{\substack{d,d\neq l\\d\neq k}}s_l^2s_ks_d\\
&=\sum_{l}\sum_{k,k\neq l} s_l^2 s_k \bigg(\sum_d s_d-s_l-s_k\bigg)\\
&=\sum_{l} s_l^2\bigg((1-s_l)\sum_{k,k\neq l}s_k-\sum_{k,k\neq l}s_k^2\bigg)\\
&=\sum_{l} s_l^2\bigg((1-s_l)^2-(1-s_l^2)\bigg)\\
&=\sum_l 2s_l^2(s_l^2-s_l)\\
&=2\left(\frac{1}{3}-\frac{1}{2}\right)=-\frac{1}{3},\\
S_2&=\sum_{l}\sum_{k,k\neq l}\sum_{\substack{d,d\neq l\\d\neq k}}\sum_{\substack{m,m\neq d\\m\neq l\\m\neq k}}s_ls_ks_ds_m\\
&=\sum_{l}\sum_{k,k\neq l}\sum_{\substack{d,d\neq l\\d\neq k}}s_ls_ks_d(1-s_d-s_l-s_k)\\
&=\sum_{l}\sum_{k,k\neq l}s_ls_k\bigg( (1-s_l-s_k)\sum_{\substack{d,d\neq l\\d\neq k}}s_d-\sum_{\substack{d,d\neq l\\d\neq k}}s_d^2\bigg)\\
&=\sum_{l}\sum_{k,k\neq l}s_ls_k\bigg( (1-s_l-s_k)^2-(1-s_l^2-s_k^2)\bigg)\\
&=\sum_{l}s_l\bigg( 2s_l(s_l-1)(1-s_l)+\sum_{k,k\neq l}2s_k(s_k^2+s_ls_k-s_k)\bigg)\\
&=\sum_{l}s_l(-6s_l^3+6s_l^2-1))\\
&=-\frac{6}{3}+\frac{6}{3}-1=0.
\end{align}

\bibliographystyle{ieeetran}
\bibliography{ref}

\end{document}